\documentstyle[11pt,titlepage]{article}
\advance\textheight by \topskip
\newlength{\bredde}
\def\slash#1{\settowidth{\bredde}{$#1$}\ifmmode\,\raisebox{.15ex}{/}
\hspace*{-\bredde} 
#1\else$\,\raisebox{.15ex}{/}\hspace*{-\bredde} #1$\fi}
\def\d{{\rm d}}
\def\e{{\rm e}}
\begin{document}
\thispagestyle{empty}
\begin{titlepage}
\addtolength{\baselineskip}{.7mm}
\thispagestyle{empty}
\begin{flushright}
ITP-UH-16/96\\
NBI-HE-96-43\\
NORDITA-96/59\\
hep-th/9609174\\
\end{flushright}
\vspace{10mm}
\begin{center}
{\Large{\bf
Universality of random matrices in the microscopic limit\\
and the Dirac operator spectrum
}}\\[15mm]
{\sc 
G.~Akemann$^{*}$,
P.~H.~Damgaard$^{\dagger}$,
U.~Magnea$^{\ddagger}$ and
S.~Nishigaki$^{\S}$} \\
\vspace{10mm}
{\it ${}^*$Institut f\"{u}r Theoretische Physik,
Universit\"{a}t Hannover\\
Appelstra\ss e 2, D-30167 Hannover, Germany \\[3mm]
${}^{\dagger\, \S}$The Niels Bohr Institute / ${}^{\ddagger}$NORDITA\\
Blegdamsvej 17, DK-2100 Copenhagen \O, Denmark}\\[6mm]
\vspace{13mm}
{\bf Abstract}\\[5mm]
\end{center}
We prove the universality of
correlation functions of chiral unitary and unitary ensembles
of random matrices in the microscopic limit. 
The essence of the proof consists in reducing the three-term recursion
relation for the relevant
orthogonal polynomials into a Bessel equation governing
the local asymptotics around the origin.
The possible physical interpretation as the universality of
soft spectrum of the Dirac operator is briefly discussed.\\

\vfill
\renewcommand{\thefootnote}{}
\footnotetext{
e-mail addresses: ${}^*$ akemann@itp.uni-hannover.de,
${}^\dagger$ phdamg, ${}^\ddagger$ umagnea, 
${}^{\S}$ shinsuke@alf.nbi.dk}
\end{titlepage}
\newpage
\renewcommand{\theequation}{\arabic{section}.\arabic{equation}}
\renewcommand{\thefootnote}{\fnsymbol{footnote}}
\setcounter{footnote}{0}
\section{Introduction}
Consider a four-dimensional gauge theory coupled to fermions
in given representations of the gauge group, and with given global
(and possibly spontaneously broken) flavor symmetries. 
In the chiral limit, the associated Dirac operator 
${\rm i} D\!\llap{/}$ anticommutes with $\gamma_5$,
\begin{equation}
\{{\rm i} D\!\llap{/},\gamma_5\} ~=~ 0 ~.
\end{equation}
This means that the associated eigenvalues of ${\rm i} D\!\llap{/}$,
defined by
${\rm i} D\!\llap{/}\,\phi_n = \lambda_n\phi_n$ 
appear symmetrically around zero:
$\pm\lambda_n$. The corresponding eigenfunctions are, for $\lambda_n
\neq 0$, $\phi_n$ and $\gamma_5\phi_n$. The accumulation of eigenvalues
near $\lambda = 0$ determines whether or not chiral symmetry is
spontaneously broken through the formation of a chiral condensate 
\cite{BC}. The study of the Dirac operator spectrum near $\lambda = 0$
is thus of fundamental importance for our understanding of chiral
symmetry breaking in gauge theories. For example, for QCD with $N_f$
flavors it is expected to lead to the conventional
${\rm SU}(N_f)_{\rm L}\times 
{\rm SU}(N_f)_{\rm R}\rightarrow{\rm SU}(N_f)_{\rm V}
$ scenario, with all its
implications in terms of low-energy effective Lagrangians.

\vspace{12pt}\noindent
In a series of papers \cite{V93,V94,V96}, 
Verbaarschot and collaborators
have added substantial new insight to this issue. Their central
assertion is that the spectral density of the Dirac operator very close
to the origin $\lambda = 0$ should be {\em universal},
depending only on the symmetries in question. One startling consequence
of this conjecture is that the spectral density of the Dirac operator
near the origin need not be computed in the gauge theory at all, but
can be extracted from much simpler random matrix theories.
The pertinent
random matrix ensemble is determined by symmetry arguments alone.
To find universal features of the Dirac spectrum near the origin,
it is essential to first consider the problem in a finite volume $V_4$,
corresponding in matrix-model terms to finite-size matrices, and then
magnify the spectrum near the origin at a scale of order $1/V_4$. 
The usual spectral density $\rho(\lambda)$ 
is defined by the following average 
over gauge field configurations:
\begin{equation}
\rho(\lambda) = \sum_n \langle\delta(\lambda - \lambda_n)\rangle ~.
\end{equation}
By the Banks-Casher relation\footnote{
Applied na\"{\i}vely here, ignoring
the subtlety of regularization.} \cite{BC}
(taking here the limit $V_4 \rightarrow \infty$),
\begin{eqnarray}
\langle \bar{\psi}\psi\rangle =\frac{1}{V_4}
\lim_{m\rightarrow 0}2m \int_0^{\infty}
\d\lambda \frac{\rho(\lambda)}{\lambda^2 + m^2} 
= \frac{\pi\rho(0)}{V_4} ~, 
\label{BC}
\end{eqnarray}
this spectral density, evaluated at the origin, is directly related to
the appearance of a chiral condensate. 
It follows that the average spacing
between eigenvalues becomes roughly constant near 
$\lambda = 0$ \cite{LS},
\begin{equation}
\Delta \lambda \sim \frac{\pi}{V_{4}\langle\bar{\psi}\psi\rangle} ~.
\end{equation}
Magnifying $\rho(\lambda)$ near
$\lambda = 0$ according to the prescription given above
thus entails the introduction of an associated 
{\em microscopic} spectral
density $\rho_{\rm S}(\lambda)$, which is defined by \cite{V93}
\begin{equation}
\rho_{\rm S}(\lambda) ~=~ \lim_{V_{4}\rightarrow\infty} \frac{1}{V_4}
\rho\left(\frac{\lambda}
{V_4}\right) ~.
\end{equation}
The precise statement is that this microscopic spectral
density should be a universal function.
The most compelling
evidence comes from the fact that the microscopic spectral
density defined as above, and evaluated in a particular random matrix
theory, exactly reproduces the Leutwyler-Smilga spectral sum rules
\cite{LS} and appropriate generalizations \cite{SV}.

\vspace{12pt}\noindent
To understand the significance of this universality conjecture, let us
for convenience first
focus on one example, that of QCD with $N_f$ massless fermions in the
fundamental representation. As
explained in refs.~\cite{V93},
the relevant matrix model partition function
is given by a chiral unitary ensemble of the form
\begin{equation}
Z_{\chi {\rm UE}} = \int 
\d W\, {\rm det}^{N_f}\left(
\begin{array}{cc}
0&W^\dagger\\
W&0
\end{array}\right)\, \e^{-N{\rm tr}\, V(W^{\dagger}W)} ~.
\end{equation}
Here only symmetry arguments alone
have selected the integration to be over complex $N\times N$
matrices, and the Dirac
operator structure is encoded in the determinant.
But such symmetry arguments
alone place no restrictions on the form of $V(\lambda)$,
which appears in the exponent. 
If indeed symmetry arguments alone should determine the 
microscopic spectral density of QCD from this matrix ensemble, then any
reasonable choice of $V(\lambda)$ should give the same result for 
$\rho_{\rm S}(\lambda)$.
This universality argument, if correct, 
allows us to evaluate the integral
with a function $V(\lambda)$ of our own choice. 
Some hints supporting this universality 
had been found in refs.~\cite{BHZ}.
Clearly, the most 
convenient choice in to make the matrix integral Gaussian except
for the determinant factor in front. This is the choice made in 
refs.~\cite{V93}.

\vspace{12pt}\noindent
It is important to separate the question of universality
of the spectral density of the Dirac operator close to 
the origin into two parts. 
The first concerns the crucial jump from the $d$-dimensional
gauge field integrations to the framework of zero-dimensional matrix 
models with similar symmetries. 
The second step concerns the conjecture about
the universality of the microscopic spectral density
{\em within the framework of matrix models}.
The purpose of this paper is to prove the latter of these
universality conjectures.
The question of whether the universality
extends all the way from zero-dimensional large-$N$ matrix models to
full-fledged quantum field theories such as QCD will of course remain
unproven. But we shall show, within the context of large-$N$ matrix
models, under what conditions the microscopic spectral density and 
higher correlation functions are universal. 
This is a major step towards
understanding the original issue, 
which concerns the possible universality
of also the Dirac operator spectrum near $\lambda = 0$.

\vspace{12pt}\noindent
This paper is organized as follows.
In section 2 we prove a theorem for the asymptotic behavior
of generic orthogonal polynomials over a semi-infinite range.
By making use of this theorem, we compute the associated
universal form of the microscopic spectral density as well as
all higher order correlation functions in the same limit.
This universality class is of relevance for four-dimensional
${\rm SU}(N_c \! \geq \! 3)$ gauge
theories with fermions in the fundamental representation.
In section 3 we repeat the analysis for a class of orthogonal
polynomials over an infinite range. This case 
is considered to be relevant
for three-dimensional ${\rm SU}(N_c \! \geq \! 3)$ 
gauge theories with an even number
of fermions in the fundamental representation. 
We conclude in section 4 with a discussion 
about the possible relation between the
chiral-flavor (or parity-flavor) symmetry breaking in gauge theories
and the now established universality of random matrices
in the microscopic limit.

\setcounter{equation}{0}
\section{Chiral unitary ensembles}

In this section we consider the chiral unitary ensemble:
\begin{equation}
Z=\int \d M \, {\rm det}^{N_f} M\, \exp 
\left\{ -\frac{N}{2} {\rm tr}\, V(M^2) \right\},\ \ \ 
V(M^2)=\sum_{k\geq 1} \frac{g_k}{k}M^{2k},\ \ \ N_f=0,1,\cdots
\label{partition} 
\end{equation}
where $M$ stands for an $(N+N')\times (N+N')$ block hermitian matrix
whose non-zero components are $N\times N'$ complex matrices
on the off-diagonals,
\begin{equation}
M=
\left(
\begin{array}{cc}
0&W^\dagger\\
W&0
\end{array}\right)
\end{equation}
and $\d M$ the Haar measure of $W$.
We take  $N\leq N'$ without loss of generality.
An ensemble is called chiral unitary because of 
the invariance under the transformation
\begin{equation}
W\mapsto V^\dagger\,W\,U ~,~~ \ \ 
U \in {\rm U}(N) ~,~~ V \in {\rm U}(N').
\label{chi-inv}
\end{equation}
This model has the same global symmetries as
a Euclidean four-dimensional ${\rm SU}(N_c\! \geq \! 3)$ gauge theory
coupled to $N_f$ massless fermions in the fundamental representation.
The topological charge $\nu$ of the vacuum is identified with
$|\nu|=N'-N$ (which is kept fixed) and 
the volume of space-time is $V_4=N+N'$ (which is sent to infinity 
afterwards).
Specifically, due to the ``${\rm U}(1)_{\rm A}$ symmetry''
\begin{equation}
\{ M, \gamma_5\}=0,\ \ 
\gamma_5=
{
\left(
\begin{array}{cc}
1_N&0\\
0&-1_{N'}
\end{array}\right)},
\end{equation}
all ($2N$) non-zero eigenvalues $M$ 
occurs in pairs with opposite signs.

\vspace{12pt}\noindent
The partition function (\ref{partition}) is expressible in terms of
the component matrices
as well as of the eigenvalues after integration 
over the angular coordinates
$(U,V)\in{\rm U}(N)\times {\rm U}(N')/{\rm U}(1)^N$ \cite{Mor},
\begin{eqnarray}
Z&=&\int 
\d W\,
{\rm det}^{N_f}\left(W^\dagger W\right)\,
\e^{-N{\rm tr} V(W^\dagger W) }
\nonumber \\
&\propto&
\int_{-\infty}^\infty \prod_{i=1}^N 
\left( \d z^2_i\, z^{2\alpha}_i\,\e^{-N V(z_i^2) }\right)
\left|\det_{ij} z^{2(i-1)}_j \right|^2\cr &=&
\int_0^\infty \prod_{i=1}^N \left( \d\lambda_i\,\lambda^{\alpha}_i\, 
\e^{-N V(\lambda_i) }\right)
\left|\det_{ij} \lambda^{i-1}_j \right|^2 ~, \label{chUE1} 
\end{eqnarray}
where we have set $\alpha=N_f+|\nu|$ and suppressed an irrelevant
constant of the angular integration.
The above expression can be interpreted as a positive definite
$N\times N$ hermitian matrix model in $H=W^\dagger W$ whose 
eigenvalues are $\lambda_1,\cdots,\lambda_N\geq 0$.
Therefore the problem is reduced to finding 
a set of orthogonal polynomials 
$P_n^{(\alpha)}(\lambda)$
over the semi-infinite interval $[0,\infty)$ with the shown
weight function,
\begin{equation}
\int_0^\infty \d\lambda\, \lambda^\alpha\,\e^{-N V(\lambda) }
 P_n^{(\alpha)}(\lambda)\,P_m^{(\alpha)}(\lambda)=
 h^{(\alpha)}_n\, \delta_{nm} .
\label{orthochUE}
\end{equation}
The definition of the orthogonality relation (\ref{orthochUE})
shows that
the problem is well-defined for arbitrary real $\alpha > -1$. 
Here we shall only deal with $\alpha$ to be a nonnegative integer,
which is motivated physically from the above.

\vspace{12pt}\noindent
In the orthogonal polynomial method,
the polynomials $P^{(\alpha)}_n(\lambda)$
are usually normalized to be monic,
\begin{equation}
P_n^{(\alpha)}(\lambda) = \lambda^n + \cdots ~,
\end{equation}
so that the Vandermonde determinant $\det_{ij} \lambda^{i-1}_j$
can be substituted by $\det_{ij} P_{i-1}^{(\alpha)}(\lambda_j)$ 
in the integrand.
However, for the present purposes 
we will employ a different normalization
by demanding $P_n^{(\alpha)}(0) = 1$:
\begin{equation}
P_n^{(\alpha)}(\lambda) = 1+ \cdots + p^{(\alpha)}_n \lambda^n.
\label{normal}
\end{equation}
This is because we shall seek a smooth limiting
behavior near $\lambda = 0$ as $n$ gets as large as
$N$ (which is taken to infinity),
and $N^2 \lambda$ is kept fixed at the same time 
(the microscopic limit). 
We thus start by the assumption that
$P_n^{(\alpha)}(0) \neq 0$, 
and normalize the polynomials accordingly.\footnote{
We avoid a proliferation of new symbols by employing the same
$P^{(\alpha)}_n(\lambda)$ and $h^{(\alpha)}_n$
in both conventions.}
As we shall see shortly, 
this assumption is equivalent to having the origin 
included in the support of the spectral density.

\subsection{Asymptotics of orthogonal polynomials}
We shall prove the following theorem on the asymptotic behavior 
of these orthogonal polynomials:
\begin{quote}
{\bf Theorem 1}:\  
{\it
Let $\{P^{(\alpha)}_n(\lambda)\}_{n=0,1,\cdots}$ be 
the set of polynomials orthogonal 
with respect to the measure} 
\[
\d \lambda\, \lambda^\alpha\,\e^{-N\,V(\lambda)},\ \ \ 
V(\lambda)=\sum_{k\geq 1} \frac{g_k}{k}\,\lambda^k,\ \ \ 
\alpha=0,1,\cdots
\]
{\it over the range $[0,\infty)$, whose moments are all finite.
If the polynomials can be normalized according to 
$P^{(\alpha)}_n(0)=1$,
then, for fixed $x=N^2\lambda$ and $t=n/N$, 
the following limiting relation holds:}
\begin{equation}
\lim_{N\rightarrow\infty} \left.
P^{(\alpha)}_{n}( \frac{x}{N^2})\right|_{n=Nt}=
\alpha !\,
 \frac{J_\alpha\left(u(t)\sqrt{x}\right)}{\left(u(t)
\sqrt{x}/2\right)^\alpha} ,
\label{Jalpha}
\end{equation}
{\it where $u(t)$ is determined by}
\[
u(t)=\int_0^t \frac{\d t'}{\sqrt{r(t')}},\ \ \ \  
t= \sum_k \frac{g_k}{2} \left(2k \atop k \right) r(t)^k.
\]
\end{quote}
The proof of this theorem for $\alpha=0$ was recently given by one of
the authors \cite{Nis}, and is reproduced here for completeness. 
The central idea of how to prove it for arbitrary
$\alpha = 1, 2, \ldots$ is to use induction, starting with the case 
$\alpha = 0$. Tracing it back, this means that the Bessel-type behavior
of the polynomials is a consequence of the three-term
recursion relation for the orthogonal polynomials.
As $n, N \rightarrow \infty$ with
$t = n/N$ fixed, this yields a certain ``continuum limit'' reminiscent
of the derivation of string equations in the double-scaling limit. The
recursion relations thus become the defining second order differential
equations for Bessel (and Neumann) functions. 
It is the boundary conditions
that finally uniquely specify the Bessel solutions. 
However, for simplicity we shall here show the proof by induction.

\vspace{12pt}\noindent
We start with the case $\alpha = 0$.
The recursion relation for 
$P_n(\lambda) \equiv P^{(0)}_n(\lambda)$ takes the form
(with $h_n \equiv h^{(0)}_n$ and $p_n \equiv p^{(0)}_n$):
\begin{eqnarray}
\lambda\,P_n(\lambda)&=&
-r_{n}\left\{ P_{n+1}(\lambda)- P_n(\lambda)
-\frac{h_{n}}{r_n}\frac{r_{n-1}}{h_{n-1}} 
\left( P_{n}(\lambda)-P_{n-1}(\lambda) 
\right) \right\} \ \ \ \ 
\left( r_n\equiv-\frac{p_{n}}{p_{n+1}}\right)
\nonumber\\
&\equiv& \sum_m \hat{\lambda}_{nm}\,P_m(\lambda). 
\label{3t-recII}
\end{eqnarray}
This is the well-known three-term recursion relation
rewritten in our nonmonic normalization. 
The structure of the coefficients is dictated by a comparison of
$O(\lambda^0)$ and $O(\lambda^{n+1})$ terms on both sides.
The sets of unknowns $\{h_n\}$, $\{r_n\}$ can  
be determined iteratively by
\begin{eqnarray}
&&1=
-\int_0^\infty 
\d\lambda \frac{\d}{\d\lambda} \left\{ 
\e^{-N\,V(\lambda)}P_n(\lambda)\,P_{n}(\lambda) \right\}=
N\,V'(\hat{\lambda})_{nn}\, h_n ,
\label{vnn}\\
&&0=
-\int_0^\infty 
\d\lambda \frac{\d}{\d\lambda} \left\{ 
\e^{-N\,V(\lambda)}\lambda\,P_{n}(\lambda)\,
P_{n}(\lambda) \right\}=
\left( N (\hat{\lambda}V'(\hat{\lambda}))_{nn}
-2n-1 \right) h_{n}.
\label{vnn-1}
\end{eqnarray}

\vspace{12pt}\noindent
In the following we need to know 
the asymptotic behavior of 
$r_n$ and $h_n$ for 
\begin{equation}
n,\ N\rightarrow \infty\ \ ~~~  
\mbox{while}~~\ \ \frac{n}{N}=t\ ~~ \mbox{is kept fixed}.
\label{lim}
\end{equation}
Eqs.~(\ref{3t-recII}), (\ref{vnn}) and (\ref{vnn-1}) 
tell us that
they should behave as
\begin{equation}
r_n=r\left(\frac{n}{N}\right)
+ \mbox{~~higher orders in ~}\frac{1}{N} ~,~~~ \ \ \ 
h_n=\frac{1}{N}\, h\left(\frac{n}{N}\right)
+ \mbox{~~higher orders in ~}\frac{1}{N}. 
\end{equation}
Then the leading behavior of the matrix $\hat{\lambda}$
and its powers is
\begin{eqnarray}
&&\hat{\lambda}_{nm}= r(\frac{n}{N})
\left(
-\delta_{n\,m-1}+2\delta_{n m}-\delta_{n\,m+1}
\right), \nonumber\\
&&(\hat{\lambda}^k)_{nm}= r(\frac{n}{N})^k
\sum_{\ell=-k}^k (-)^\ell \left( {2k \atop k+\ell} \right)
\delta_{n\,m+\ell}
\end{eqnarray}
so that eqs.~(\ref{vnn}) and (\ref{vnn-1}) read
\begin{eqnarray}
\sum_{k} g_k \left({2k-2\atop k-1}\right) 
r(t)^{k-1}
&=&
\frac{1}{h(t)} ,
\label{vNN}\\
\frac12 \sum_{k} g_k \left({2k \atop k}\right) 
r(t)^{k} 
&=&t.
\label{qn}
\end{eqnarray}
Note that eqs.~(\ref{vNN}) and (\ref{qn}) 
imply a universal relationship
among total derivatives,
\begin{equation}
\d t=
2r\,\d\left( \frac1h \right) + \frac{1}{h} \d r
=\sqrt{r}\, \d \left( \frac{2\sqrt{r}}{h} \right).
\label{dt}
\end{equation}

\vspace{12pt}\noindent
Next we expand the right hand side of the recursion relation 
(\ref{3t-recII}) in terms of $1/N$ in the limit (\ref{lim}) to get
\begin{equation}
N^2 \lambda \, P(n,N,\lambda) =
-r(t)\left\{\frac{\d^2\,P}{\d t^2}+
\frac{h(t)}{r(t)}
 \left( \frac{\d}{\d t}  \frac{r(t)}{h(t)}\right) \frac{\d P}{\d t}
\right\}
=
-h(t) \frac{\d}{\d t}  \frac{r(t)}{h(t)} \frac{\d}{\d t}
 P(n,N,\lambda) ,
\label{diff-eq}
\end{equation}
where the argument $N$ in 
$P(n, N, \lambda)\equiv P_n(\lambda)$ is to indicate 
explicitly the dependency via the coefficient 
in front of the potential.
Only the leading terms up to $O(1/N)$ have been kept here.
Eq.~(\ref{diff-eq}) tells us that that the arguments of the polynomial
appear
only in the combinations $t=n/N$ and $x=N^2\lambda$
in the large-$N$ limit.
The rescaled eigenvalue coordinate $x$ is to be fixed finite hereafter,
and is regarded as a parameter 
in the ordinary differential equation in $t$.
Performing the change of variable
\begin{equation}
t \mapsto u(t)\equiv \frac{2\sqrt{r(t)}}{h(t)}=
\int_0^t \frac{\d t'}{\sqrt{r(t')}},
\end{equation}
using the relationship (\ref{dt}),
and neglecting higher order terms in $1/N$,
eq.~(\ref{diff-eq}) 
reduces to Bessel equation of zeroth order:
\begin{equation}
\left(
\frac1u \frac{\d}{\d u}  u \frac{\d}{\d u} +x
\right)
P(u,x)=0.
\end{equation}
The unique solution to this equation which satisfies
the following boundary condition at $t=n/N=0$ ($u(0)=0$),
\begin{equation}
P(0, x)=P_0(\lambda)=1 ,
\end{equation}
is
\begin{equation}
P(u,x)=J_0\left( u\sqrt{x} \right) .
\label{Bes0}
\end{equation}

\vspace{12pt}\noindent
We now proceed for a generic integer $\alpha$ by induction.
Given a set of polynomials $\{P^{(\alpha)}_n(\lambda)\}$
which are orthogonal with respect to the measure 
$\d\lambda \,\lambda^\alpha\,\e^{-NV(\lambda)}$
on $[0,\infty)$,
and are normalized by $P^{(\alpha)}_n(0)=1$ as in eq.~(\ref{normal}), 
we note that
\begin{equation}
\tilde{P}_n^{(\alpha+1)}(\lambda) \equiv 
\frac{P^{(\alpha)}_{n+1}(\lambda)-P^{(\alpha)}_n(\lambda)}{\lambda}
\label{Ptn+1}
\end{equation}
are polynomials of order $n$. They are orthogonal to
$\lambda^k$, 
$k=0,\cdots,n-1$
with respect to the measure 
$\d\lambda\,\lambda^{\alpha+1}\,\e^{-NV(\lambda)}$
\cite{Sze}:
\begin{eqnarray}
&&\int_0^\infty \d\lambda\,\lambda^{\alpha+1}\,\e^{-NV(\lambda)}
\tilde{P}_n^{(\alpha+1)}(\lambda) \,\lambda^k
\nonumber\\
&=&\int_0^\infty \d\lambda\,\lambda^{\alpha}\,\e^{-NV(\lambda)}
\left( P^{(\alpha)}_{n+1}(\lambda)-P^{(\alpha)}_n(\lambda) \right)
\lambda^k =0.
\end{eqnarray}
Under the assumption that $\tilde{P}_n^{(\alpha+1)}(0)\neq 0$,
the orthogonal polynomial normalized to ${P}_n^{(\alpha+1)}(0) =1$
is thus given by
\begin{equation}
P^{(\alpha+1)}_n(\lambda) =
\frac{\tilde{P}_n^{(\alpha+1)}(\lambda) }{\tilde{P}_n^{(\alpha+1)}(0)}.
\label{Pn+1}
\end{equation}
The microscopic limit of the contiguous relation (\ref{Ptn+1}) reads
\begin{equation}
\tilde{P}^{(\alpha+1)}_{Nt} (\frac{x}{N^2})=
\frac{N}{x}\frac{\partial}{\partial t} 
P^{(\alpha)}_{Nt} (\frac{x}{N^2})
+ \mbox{~~higher orders in ~}\frac{1}{N}.
\end{equation}
If we substitute eq.~(\ref{Jalpha}) 
for a given $\alpha$ into the above, we confirm that
\begin{eqnarray}
\tilde{P}^{(\alpha+1)}_{Nt} (\frac{x}{N^2})&=&
-N\, \frac{\alpha!}{\sqrt{r(t)}}
\frac{
J_{\alpha+1}\left(u(t)\sqrt{x}\right)}{\left(u(t)/2\right)^\alpha
(\sqrt{x})^{\alpha+1}},
\\
\tilde{P}^{(\alpha+1)}_{Nt} (0)&=&
-N\,\frac{u(t)/2}{(\alpha+1)\sqrt{r(t)}}.
\end{eqnarray}
Thus eq.~(\ref{Jalpha}) holds also for $\alpha+1$ 
after normalizing according to (\ref{Pn+1}).
Together with the result (\ref{Bes0}) for $\alpha=0$,
this completes the proof.

\vspace{12pt}\noindent
We have worked out 
an alternative proof of the same
theorem, which does not rely on induction. 
Rather, it is possible to convert the recursion relation for 
$P^{(\alpha)}_n(\lambda)$
directly into Bessel equation of order $\alpha$,
$\left( \frac{1}{u^{2\alpha+1}}\frac{\d}{\d u}
u^{2\alpha+1}\frac{\d}{\d u}
+x\right)P^{(\alpha)}(u,x)=0$.
Since it is lengthy, we do not reproduce it here.
   
\subsection{Universal correlations}
\noindent
Theorem 1 can be used to establish a remarkable universal form of
spectral correlators of the model (\ref{chUE1}).
We shall consider
the spectral density $\rho(\lambda)$, and 
higher order correlators, in the
same microscopic limit as above.

\vspace{12pt}\noindent
To begin, we recall the expression for 
the integration kernel $K_N(\lambda,\mu)$ 
associated with the eigenvalue problem
for the positive-definite hermitian
matrix $H=W^\dagger W$,
\begin{eqnarray}
K_N(\lambda,\mu)\!&\!=\!&\!
(\lambda\,\mu)^{\alpha/2}\,
\e^{-\frac{N}{2}(V(\lambda)+V(\mu))} \,
\frac1N \sum_{i=0}^{N-1} 
\frac{P^{(\alpha)}_i(\lambda)P^{(\alpha)}_i(\mu)}{h^{(\alpha)}_i}
\nonumber \\
\!&\!=\!&\!
(\lambda\,\mu)^{\alpha/2}\,
\e^{-\frac{N}{2}(V(\lambda)+V(\mu))} \,
\frac1N\,\frac{-r^{(\alpha)}_{N-1}}{h^{(\alpha)}_{N-1}}
\frac{P^{(\alpha)}_N(\lambda)P^{(\alpha)}_{N-1}(\mu)-
P^{(\alpha)}_{N-1}(\lambda)P^{(\alpha)}_{N}(\mu)}{\lambda-\mu}.
\label{kernel}
\end{eqnarray}
Here use has been made of the Christoffel-Darboux identity.
In the large-$N$ limit
we may drop the superscript $\alpha$ in 
$r^{(\alpha)}_n=-p^{(\alpha)}_n/p^{(\alpha)}_{n+1}$. This is 
because the general recursion relation 
(the counterpart of eq. (\ref{vnn-1})
for general $\alpha$)
\begin{equation}
0=
N (\hat{\lambda}V'(\hat{\lambda}))_{nn}
-2n-\alpha-1
\label{vnn-1alpha}
\end{equation}
which determines $r^{(\alpha)}(t)$
leads, in the large-$N$ limit (\ref{lim}), 
to the same (\ref{qn}) for any finite $\alpha$.

\vspace{12pt}\noindent
On the other hand, the norms of the polynomials can be evaluated 
iteratively:
\begin{eqnarray}
\tilde{h}^{(\alpha+1)}_n&=&
\int_0^\infty \d\lambda\,\lambda^{\alpha+1}\,\e^{-NV(\lambda)}
\tilde{P}_n^{(\alpha+1)}(\lambda) 
\tilde{P}_n^{(\alpha+1)}(\lambda) 
\nonumber\\
&=&\int_0^\infty \d\lambda\,\lambda^{\alpha}\,\e^{-NV(\lambda)}
\left( P^{(\alpha)}_{n+1}(\lambda)-P^{(\alpha)}_n(\lambda) \right)
\tilde{P}_n^{(\alpha+1)}(\lambda) 
\nonumber\\
&=&\int_0^\infty \d\lambda\,\lambda^{\alpha}\,\e^{-NV(\lambda)}
\left(-P^{(\alpha)}_n(\lambda) \right)
\frac{p^{(\alpha)}_{n+1}}{p^{(\alpha)}_n}{P}_n^{(\alpha)}(\lambda) 
=\frac{{h}^{(\alpha)}_n}{r_n} ~,\\
{h}^{(\alpha+1)}_n&=&
\frac{\tilde{h}^{(\alpha+1)}_n}{\tilde{P}^{(\alpha+1)}_n(0)^2}
={h}^{(\alpha)}_n
\left( \frac{\alpha+1}{N\,u(t)/2} \right)^2 .
\end{eqnarray}
We therefore obtain
\begin{equation}
{h}^{(\alpha)}_n =
\frac{1}{N^{2\alpha+1}}
(\alpha!)^2
\left(\frac{2}{u(t)}\right)^{2\alpha} h(t)
+\mbox{~higher orders in }\frac1N
\ \ \ \ \left(h(t) \equiv h^{(0)}(t) \right) ~.
\label{halpha}
\end{equation}
Using Theorem 1 and inserting eq.~(\ref{halpha}) into 
(\ref{kernel}),
we obtain a universal form of the kernel (the Bessel kernel
of order $\alpha$)
in the microscopic limit
\begin{equation}
\lim_{N\rightarrow\infty} \frac1N\, 
K_N(\frac{x}{N^2}, \frac{y}{N^2})=
\frac{u(1)}{2} 
\frac{
\sqrt{x} J_{\alpha+1} ( u(1)\sqrt{x} ) J_\alpha ( u(1)\sqrt{y} )
-
J_\alpha ( u(1)\sqrt{x} ) \sqrt{y} J_{\alpha+1} ( u(1)\sqrt{y} )
}{x-y} .
\label{Besker}
\end{equation}

\vspace{12pt}\noindent
The only way this asymptotic kernel depends on the potential
$V(\lambda)$ 
is through the parameter $u(1)$.
As we shall show now, even this dependence
is of a highly universal form. 
Let us compare eqs.~(\ref{vNN}) and (\ref{qn}) at $t=1$
with the explicit expression for the
large-$N$ spectral density $\rho(z)$ 
having  support on a single interval $[-a,a]$
\cite{BB}. 
\begin{eqnarray}
&&\rho(z)=\frac{\sqrt{a^2-z^2}}{2\pi} 
\sum_{k} g_k \sum_{n=0}^{k-1}
\left( {2n \atop n} \right) \left({a^2 \over 4}\right)^n
z^{2k-2n-2},\\
&& \frac12 \sum_{k} g_k \left({2k \atop k}\right) 
\left( {a^2 \over 4}\right)^{k}=1 ~.
\end{eqnarray}
This enables us to relate the parameters $r(1)$ and $u(1)$ to
$a$ and $\rho(0)$, respectively,
\begin{equation}
a ~=~ 2\sqrt{r(1)} ~~~,~~~~\ \ 
\rho(z\!\!=\!\!0) ~=~ \frac{u(1)}{2\pi}.
\label{rho0}
\end{equation}
The kernel (\ref{Besker}) 
therefore only depends on the potential $V(\lambda)$
in the indirect way of setting the scale $\rho(0)$. It has clearly been
the assumption throughout that $\rho(\lambda) \neq 0$. This was the
starting point of the physical motivation from field theory
(spontaneously
broken chiral symmetries, which through the Banks-Casher relation 
(\ref{BC}) entails a non-vanishing spectral density at the origin).
Moreover, it is known that the critical condition 
$\rho(0)=0$
for a transition where the intervals of support move away from the
origin is equivalent to $P_N(0)=0$ 
for the monically normalized polynomials \cite{Mor}.
If we go back to our 
proof of Theorem 1, this was precisely the condition we had to impose
in order to be able to find a smooth continuum limit of the orthogonal
polynomials in the large-$N$ limit.
Thus, under the assumption 
that the normalization (\ref{normal}) is possible,
the constants $u(1)$ and $r(1)$
are determined to be positive from
the identification (\ref{rho0}).

\vspace{12pt}\noindent
It finally remains us to relate the $s$-point correlation functions of
$\sigma_N$ ($\rho_N$) of eigenvalues of $H=W^\dagger W$ (or $M$)
to the kernel $K_N$ in eq.~(\ref{Besker}). The former is defined as
\begin{eqnarray}
&&\sigma_N(\lambda_1,\cdots,\lambda_s)=
\left\langle \prod_{a=1}^s \frac{1}{N} 
{\rm tr}\, \delta (\lambda_a-H)\right\rangle=
\det_{1\leq a,b \leq s} K_N (\lambda_a,\lambda_b),\\
&&\rho_N(z_1,\cdots,z_s)=
\left\langle \prod_{a=1}^s \frac{1}{2N} 
{\rm tr}\, \delta (z_a-M)\right\rangle=
| z_1 | \cdots | z_s |\, 
\sigma_N(z_1^2,\cdots,z_s^2),
\end{eqnarray}
respectively.
Therefore all the formulae for their 
microscopic limits ($x_a=N^2\lambda_a$, $\zeta_a=N z_a$ fixed)
\begin{eqnarray}
&&\sigma_{\rm S}(x_1,\cdots,x_s)\equiv
\lim_{N\rightarrow\infty}\frac{1}{N^s}
\sigma_N (\frac{x_1}{N^2},\cdots,\frac{x_s}{N^2}) ,\\
&&\rho_{\rm S}(\zeta_1,\cdots,\zeta_s)\equiv
\lim_{N\rightarrow\infty}
\rho_N(\frac{\zeta_1}{N},\cdots,\frac{\zeta_s}{N})=
| \zeta_1 | \cdots | \zeta_s |\, 
\sigma_{\rm S}(\zeta_1^2,\cdots,\zeta_s^2),
\end{eqnarray}
that were previously calculated for 
the Laguerre (in the $H$-picture) or 
chiral Gaussian (in the $M$-picture) 
unitary ensemble \cite{V93,NS}, hold universally.

\vspace{12pt}\noindent
In particular, 
the spectral density of the chiral unitary ensemble
\begin{equation}
\rho_N(z)=\left\langle \frac{1}{2N}
\, {\rm tr}\, \delta( z- M
)\right\rangle 
=| z | K_N (z^2, z^2)
\end{equation}
takes the universal form
\begin{equation}
\rho_{\rm S} (\zeta)=
\left( \pi \rho(0) \right)^2
|\zeta|
\left(
J_\alpha^2(2\pi\rho(0)  \zeta) -
J_{\alpha+1}(2\pi\rho(0)  \zeta)
J_{\alpha-1}(2\pi\rho(0)  \zeta)
\right)
\end{equation}
in the microscopic limit.
As expected from the Gaussian case, 
a matching condition
between the microscopic and macroscopic (ordinary large-$N$) 
spectral densities is satisfied:
\begin{equation}
\lim_{\zeta\rightarrow\infty}\rho_{\rm S}(\zeta)=
\rho(0) .
\end{equation}

\setcounter{equation}{0}
\section{Unitary ensembles}
In this section we consider the unitary ensemble:
\begin{equation}
Z=\int \d M\,
{\rm det}^{2\alpha}M\,
 \e^{-N{\rm tr} V(M^2) } ,
\ \ \ 
V(M^2)=\sum_{k\geq 1} \frac{g_k}{2k}M^{2k},
\ \ \ \alpha=0,1,\cdots
\label{partition2}
\end{equation}
where $M$ stands for an $N\times N$ hermitian matrix and
$\d M$ its Haar measure.
This model shares the same global symmetries with
a Euclidean three-dimensional ${\rm SU}(N_c\! \geq \! 3)$ gauge theory
coupled to $N_f=2\alpha$ massless fermions 
in the fundamental representation
\cite{VZ}.
The partition function (\ref{partition2}) is expressible in terms of
the eigenvalues after integration over 
the angular ${\rm U}(N)$ coordinates
\begin{equation}
Z=\int_{-\infty}^\infty \prod_{i=1}^N 
\left( \d\lambda_i\ \lambda_i^{2\alpha}\, 
\e^{-N V(\lambda_i^2) }\right)
\left|\det_{i,j}\lambda^{j-1}_i \right|^2 .
\end{equation}
Therefore the problem is reduced to finding 
a set of orthogonal polynomials 
$P^{(\alpha)}_n(\lambda)$
over the infinite interval $(-\infty,\infty)$ with the shown
weight function,
\begin{equation}
\int_{-\infty}^\infty \d \lambda\, 
\lambda^{2\alpha}\,\e^{-N V(\lambda^2) }
 P^{(\alpha)}_n(\lambda)\,P^{(\alpha)}_m(\lambda)=
 h^{(\alpha)}_n\, \delta_{nm}.
\label{orthoUE}
\end{equation}

\subsection{Asymptotics of orthogonal polynomials}
We shall prove the following theorem\footnote{
Theorem 2 can be regarded as an extension of
Theorem 1 for half-integer $\alpha$.}
on the asymptotic behavior 
of these orthogonal polynomials:\
\newcounter{sub}
\setcounter{sub}{1}
\renewcommand{\theequation}{%
\arabic{section}.\arabic{equation}\alph{sub}}
\begin{quote}
{\bf Theorem 2}:\  
{\it Let $\{P^{(\alpha)}_n(\lambda)\}_{n=0,1,\cdots}$ be 
the set of polynomials orthogonal
with respect to the measure
\[
\d \lambda\, \lambda^{2\alpha}\,\e^{-NV(\lambda^2)},\ \
V(\lambda^2)=\sum_{k\geq 1} \frac{g_k}{2k}\,\lambda^{2k},\ \
\alpha=0,1,\cdots
\]
over the range $(-\infty,\infty)$,
whose moments are all finite.
If the polynomials can be normalized according to
$P^{(\alpha)}_{2n}(0)=P^{(\alpha)}_{2n+1}{}'(0)=1$,
then, for fixed $x=N\,\lambda$ and $t=2n/N$, 
the following limiting relations
hold:}
\begin{eqnarray}
&&\lim_{N\rightarrow\infty}\left.
P^{(\alpha)}_{2n}( \frac{x}{N})\right|_{n=Nt/2}=
\Gamma(\alpha+\frac12)\,
\frac{J_{\alpha-\frac12}
\left(u(t)x\right)}{\left(u(t)x/2\right)^{\alpha-\frac12}} ,
\label{Jalphaminushalf} \\
\addtocounter{equation}{-1}
\addtocounter{sub}{1}
&&\lim_{N\rightarrow\infty}\left.
{N}\, P^{(\alpha)}_{2n+1}( \frac{x}{N})\right|_{n=Nt/2}=
x\,
\Gamma(\alpha+\frac32)\,
\frac{J_{\alpha+\frac12}
\left(u(t)x\right)}{\left(u(t)x/2\right)^{\alpha+\frac12}} ,
\label{Jalphahalf}
\end{eqnarray}
{\it where $u(t)$ is determined by}
\[
u(t)=\int_0^t \frac{\d t'}{2\sqrt{r(t')}},\ \ \ \
t= \sum_k \frac{g_k}{2} \left(2k \atop k \right) r(t)^k.
\]
\end{quote}
\renewcommand{\theequation}{\arabic{section}.\arabic{equation}}
The proof of this theorem for $\alpha=0$ was sketched
in ref.~\cite{Moo}, and below we shall elaborate it in a rigorous form.
The recursion relation for monically normalized polynomial
\begin{equation}
P_n(\lambda)\equiv P^{(0)}_n(\lambda)=\lambda^n+\cdots
\label{monic}
\end{equation}
takes the form
\begin{equation}
\lambda\,P_n(\lambda)=
P_{n+1}(\lambda)+ r_n\,P_{n-1}(\lambda),
\label{rec}
\end{equation}
with 
\begin{equation}
r_n= \frac{h_n}{h_{n-1}} ~.
\end{equation}
Since the monic normalization (\ref{monic})
does not distinguish
the parity of $n$, neither does $r_n$.
Eq.~(\ref{rec}) immediately implies
\begin{equation}
0=
P_{2n+2}(0)+ r_{2n+1}\,P_{2n}(0).
\label{rn}
\end{equation}
Due to the ${\bf Z}_2$ symmetry of the weight function,
the polynomials have definite parities.
Therefore we are lead to change the normalization of polynomials 
as follows:
\begin{equation}
\frac{P_{2n}(\lambda)}{P_{2n}(0)}
\rightarrow P_{2n}(\lambda)=1+\cdots, \ \ \ 
\frac{P_{2n+1}(\lambda)}{P_{2n+1}{}'(0)}
\rightarrow P_{2n+1}(\lambda)=\lambda+\cdots.
\end{equation}
Then the recursion relation (\ref{rec}) in this normalization,
applied twice to $P_{2n}(\lambda)$, takes the form
\begin{eqnarray}
\lambda^2\,P_{2n}(\lambda)&=&
-r_{2n+1}\left( P_{2n+2}(\lambda)- P_{2n}(\lambda)\right)
+r_{2n}\left( P_{2n}(\lambda)- P_{2n-2}(\lambda)\right)\nonumber\\
&\equiv& \sum_m (\hat{\lambda}^2)_{2n,2m} \,P_{2m}(\lambda).
\label{receven}
\end{eqnarray}
Here use is made of eq.~(\ref{rn}).
A similar equation for $P_{2n+1}(\lambda)$ expresses the matrix 
elements $(\hat{\lambda}^2)_{2n+1,2m+1}$ in terms of the $r_n$'s.
The coefficients $\{r_n\}$ are iteratively determined by
\begin{equation}
0=
-\frac{1}{h_n} \int_{-\infty}^\infty 
\d\lambda \frac{\d}{\d\lambda} \left\{ 
\e^{-N\,V(\lambda^2)}\lambda\,P_{n}(\lambda)\,
P_{n}(\lambda) \right\}=
2N (\hat{\lambda}^2 V'(\hat{\lambda}^2))_{nn}
-2n-1 .
\label{vnnII}
\end{equation}
The asymptotic behavior of $r_n$
is determined in the large-$N$ limit,
\begin{equation}
n,\ N\rightarrow \infty\ \ ~~~  
\mbox{while}~~\ \ \frac{2n}{N}=t\ ~~ \mbox{is kept fixed},
\label{limII}
\end{equation}
to be
\begin{equation}
r_{n}=r\left( \frac{n}{N} \right)
+ \mbox{higher orders in }\frac{1}{N}.
\end{equation}
Here $r(t)$ is given by
\begin{equation}
\frac12 \sum_{k} g_k \left({2k \atop k}\right) 
r(t)^{k} =t.
\end{equation}

\vspace{12pt}\noindent
Next we expand the right hand side of the 
recursion relation (\ref{receven}) in terms of $1/N$ to get
($P(2n,N,\lambda)\equiv P_{2n}(\lambda)$)
\begin{equation}
N^2 \lambda^2\, P(2n,N,\lambda)=-
\left( 4r(t)\frac{\d^2 P}{\d t^2} +
2 \frac{\d r(t)}{\d t} 
\frac{\d P}{\d t} \right)=
- \left(2\sqrt{r(t)} \frac{\d}{\d t}\right)^2 
P(2n,N,\lambda).
\end{equation}
It tells us that that the arguments of $P$ appear
only in the combinations $t=2n/N$ and $x=N\lambda$
in the large-$N$ limit.
After the change of variables
\begin{equation}
t \mapsto u(t)\equiv \int_0^t \frac{\d t'}{2\sqrt{r(t')}},
\end{equation}
we equivalently have
\begin{equation}
\left( \frac{\d^2}{\d u^2}  +x^2 \right)
P(u,x)=0.
\end{equation}
The unique solution to this trigonometric differential equation
which satisfies the boundary and parity conditions
\begin{equation}
P(0,x)=1,\ \ 
P(u,-x)=P(u,x)
\end{equation}
is
\begin{equation}
P(u,x)=
\lim_{N\rightarrow\infty}\left.\,
P_{2n}(\frac{x}{N})\right|_{n=Nt/2}=
\cos u x 
=\Gamma(\frac12) 
\frac{J_{-\frac12}\left( u x \right) }{
\left( u x/2 \right)^{-\frac12}}.
\label{J-1/2}
\end{equation}

\vspace{12pt}\noindent
The odd-order polynomials are constructed out of
normalized even-order ones as follows.
\begin{equation}
\tilde{P}_{2n+1}(\lambda) \equiv 
\frac{P_{2n+2}(\lambda)-P_{2n}(\lambda)}{\lambda}
\label{Pt2n+1}
\end{equation}
are odd polynomials of order $2n+1$, and are orthogonal to
$\lambda^{2k+1}$, $k=0,\cdots,n-1$:
\begin{eqnarray}
&&\int_{-\infty}^\infty \d\lambda\,\e^{-NV(\lambda)}
\tilde{P}_{2n+1} (\lambda) \,\lambda^{2k+1}\nonumber \\
&=&\int_{-\infty}^\infty \d\lambda\,\e^{-NV(\lambda)}
\left( P_{2n+2}(\lambda)-P_{2n}(\lambda) \right)
\lambda^{2k}
=0.
\end{eqnarray}
The microscopic limit of (\ref{Pt2n+1}) reads
\begin{equation}
\left. \tilde{P}_{2n+1} (\frac{x}{N}) \right|_{n=Nt/2} =
\frac{2}{x}\frac{\partial}{\partial t} 
\left.
P_{2n} (\frac{x}{N})\right|_{n=Nt/2} .
\label{Pt2n+1mic}
\end{equation}
Under the assumption that $\tilde{P}_{2n+1}{}'(0)\neq 0$,
change of normalization to $P_{2n+1}{}'(0)=1$
is implemented by
\begin{equation}
P_{2n+1}(\lambda) =
\frac{\tilde{P}_{2n+1}(\lambda) }{\tilde{P}_{2n+1}{}'(0) }.
\label{P2n+1}
\end{equation}
By inserting the asymptotic form of even-order polynomials
(\ref{J-1/2}) into (\ref{Pt2n+1mic}) and 
normalizing according to (\ref{P2n+1}),
we conclude
\begin{equation}
\lim_{N\rightarrow\infty}\left.N\,
P_{2n+1}(\frac{x}{N})\right|_{n=Nt/2}
=\frac{\sin u x }{u}
=x\,\Gamma(\frac32) 
\frac{J_{\frac12}\left( u x \right)}{\left( u x/2 \right)^{\frac12}}.
\label{J+1/2}
\end{equation}

\vspace{12pt}\noindent
Now we proceed for a generic $\alpha$ by induction.
Given a set of polynomials $\{P^{(\alpha)}_n(\lambda)\}$
which are orthogonal with respect to the measure 
$\d\lambda\,\lambda^{2\alpha}\,\e^{-NV(\lambda^2)}$
on $[-\infty,\infty]$, and are normalized by
$P^{(\alpha)}_{2n}(0)=P^{(\alpha)}_{2n+1}{}'(0)=1$, 
we note that 
\begin{equation}
\tilde{P}_{n}^{(\alpha+1)}(\lambda) \equiv 
\frac{P^{(\alpha)}_{n+2}(\lambda)-P^{(\alpha)}_{n}(\lambda)}{\lambda^2}
\label{Ptn+1II}
\end{equation}
are polynomials of order $n$. They are orthogonal to
each other
with respect to the measure \linebreak[3]
$\d\lambda\,\lambda^{2(\alpha+1)}\,
\e^{-NV(\lambda^2)}$.
Under the assumption that 
$\tilde{P}_{2n}^{(\alpha+1)}(0),
\tilde{P}_{2n+1}^{(\alpha+1)}{}'(0)
\neq 0$,
the correctly normalized orthogonal polynomials 
are given by
\begin{equation}
P^{(\alpha+1)}_{2n}(\lambda) =
\frac{\tilde{P}_{2n}^{(\alpha+1)}(\lambda) }{
\tilde{P}_{2n}^{(\alpha+1)}(0) }
, \ \ 
P^{(\alpha+1)}_{2n+1}(\lambda) =
\frac{
\tilde{P}_{2n+1}^{(\alpha+1)}(\lambda)}{
\tilde{P}_{2n+1}^{(\alpha+1)}{}'(0)}.
\label{n}
\end{equation}
If we substitute eqs.~(3.4)
for a given $\alpha$ into the microscopic limit of eq.~(\ref{Ptn+1II}),
we confirm that
\setcounter{sub}{1}
\renewcommand{\theequation}{%
\arabic{section}.\arabic{equation}\alph{sub}}
\begin{eqnarray}
\left. \tilde{P}^{(\alpha+1)}_{2n} (\frac{x}{N}) \right|_{n=Nt/2} &=&
-N\, \frac{\Gamma(\alpha+\frac12)}{\sqrt{r(t)}}
\frac{
J_{\alpha+\frac12}\left(u(t)x\right)
}{
\left(u(t)/2\right)^{\alpha-\frac12}
x^{\alpha+\frac12}} ,
\\
\addtocounter{equation}{-1}
\addtocounter{sub}{1}
\left. \tilde{P}^{(\alpha+1)}_{2n+1} (\frac{x}{N})\right|_{n=Nt/2}&=&
-\frac{\Gamma(\alpha+\frac32)}{\sqrt{r(t)}}
\frac{
J_{\alpha+\frac32}\left(u(t)x\right)
}{
\left(u(t)x/2\right)^{\alpha+\frac12}},
\\
\addtocounter{sub}{-1}
\left. \tilde{P}^{(\alpha+1)}_{2n} (0)\right|_{n=Nt/2}&=&
-N\,\frac{u(t)}{(2\alpha+1)\sqrt{r(t)}} ,
\\
\addtocounter{equation}{-1}
\addtocounter{sub}{1}
\left. \tilde{P}^{(\alpha+1)}_{2n+1} {}'(0)\right|_{n=Nt/2}&=&
-N\,\frac{u(t)}{(2\alpha+3)\sqrt{r(t)}}.
\end{eqnarray}
Thus eqs.~(3.4) hold also for $\alpha+1$ after normalizing according to 
(\ref{n}).
Together with the results (\ref{J-1/2}) and (\ref{J+1/2})
for $\alpha=0$, this completes the proof of Theorem 2.

\renewcommand{\theequation}{\arabic{section}.\arabic{equation}}
\subsection{Universal correlations}
The integration kernel associated with
the eigenvalue problem of $M$ is
(for brevity we set $N=\mbox{odd}=2{\cal N}+1$)
\begin{eqnarray}
K_N(\lambda,\mu)
\!\!&\!\!\!=\!\!\!&\!\!
(\lambda\,\mu)^{\alpha}\,
\e^{-\frac{N}{2}(V(\lambda^2)+V(\mu^2))} \,
\frac1N \sum_{n=0}^{N-1} 
\frac{P^{(\alpha)}_n(\lambda)P^{(\alpha)}_n(\mu)}{h^{(\alpha)}_n}
\nonumber \\
\!\!&\!\!\!=\!\!\!&\!\!
(\lambda\,\mu)^{\alpha}\,
\e^{-\frac{N}{2}(V(\lambda^2)+V(\mu^2))} 
\frac1N \!
\left( 
\sum_{n=0}^{\cal N} \frac{1}{ h^{(\alpha)}_{2n}}\right) \!
\frac{P^{(\alpha)}_{2{\cal N}+1}(\lambda)P^{(\alpha)}_{2{\cal N}}(\mu)-
P^{(\alpha)}_{2{\cal N}}(\lambda)
P^{(\alpha)}_{2{\cal N}+1}(\mu)}{\lambda-\mu}.
\label{kernelII}
\end{eqnarray}
Here use is made of the Christoffel-Darboux identity, 
and the proportionality constant in the second line is
determined by matching $O(\lambda^0 \mu^0)$ terms.
Noting that
the orthogonality relation (\ref{orthoUE}) in the microscopic limit
\begin{equation}
\frac{2}{N^{2\alpha+1}}
\int_0^\infty \d x \,x^{2\alpha}
\left. P^{(\alpha)}_{2n}(\frac{x}{N})\right|_{n=Nt/2} 
\left. P^{(\alpha)}_{2m}(\frac{x}{N})\right|_{m=Nt'/2}
= h^{(\alpha)}_{2n}\,\frac2N \delta (t-t')
\end{equation}
is identical to the Bessel closure relation
(inversion of Hankel transform)
\begin{equation}
\int_0^\infty \d x\,x\,J_{\alpha-\frac12}(ux) J_{\alpha-\frac12}(u'x) 
= \frac1u \delta (u-u'),
\end{equation}
we can determine the asymptotic form of the norm as
\begin{equation}
{h}^{(\alpha)}_{2n} = \frac{1}{N^{2\alpha}}{\Gamma(\alpha+\frac12)^2} 
\left(\frac{2}{u(t)}  \right)^{2\alpha} \sqrt{r(t)}
+\mbox{higher orders in }\frac1N.
\end{equation}
Accordingly we obtain
\begin{equation}
\sum_{n=0}^{\cal N}
\frac{1}{h^{(\alpha)}_{2n}}=
\frac{N^{2\alpha}}{\Gamma(\alpha+\frac12)^2\,2^{2\alpha}} 
\cdot \frac{N}{2}
\int_0^1 \d t \frac{u(t)^{2\alpha}}{\sqrt{r(t)}}
=\frac{N^{2\alpha+1}}{\Gamma(\alpha+\frac12)^2 \,
2^{2\alpha}}
\frac{u(1)^{2\alpha+1}}{2\alpha+1}.
\label{halpha-1/2}
\end{equation}
Using Theorem 2 and inserting eq.~(\ref{halpha-1/2}) into 
(\ref{kernelII}),
we obtain the universal form of the kernel 
(the generalized sine kernel of order $\alpha$)
in the microscopic limit
\begin{equation}
\lim_{N\rightarrow\infty}\frac1N K_N(\frac{x}{N},\frac{y}{N})=
\frac{u(1)}{2}\sqrt{xy} 
\frac{J_{\alpha+\frac12}(u(1) x)J_{\alpha-\frac12}(u(1) y)-
      J_{\alpha-\frac12}(u(1) x)J_{\alpha+\frac12}(u(1) y)}{x-y} .
\end{equation}
Therefore all formulae for the correlation functions
\begin{equation}
\rho_N(\lambda_1,\cdots,\lambda_s)=
\left\langle \prod_{a=1}^s \frac{1}{N} 
{\rm tr}\, \delta (\lambda_a-M)\right\rangle=
\det_{1\leq a,b \leq s} K_N (\lambda_a,\lambda_b) ,
\end{equation} 
previously calculated for the Gaussian unitary ensemble 
in the microscopic limit \cite{Wig,NS}
\begin{equation}
\rho_{\rm S} (x_1,\cdots,x_s)=\lim_{N\rightarrow\infty}
\frac{1}{N^s}\rho_N(\frac{x_1}{N},\cdots,\frac{x_s}{N}),
\end{equation}
hold universally.
Specifically, the spectral density
$
\rho_N(\lambda)=
\langle 1/N \,{\rm tr}\, \delta(\lambda-M)\rangle =K_N(\lambda,\lambda)
$
universally takes the form
(now $\rho(0)=u(1)/\pi$)
\begin{equation}
\rho_{\rm S} (x)=\left(\frac{\pi\rho(0)}{2}\right)^2 x \left(  
J_{\alpha+\frac12}^2 + J_{\alpha-\frac12}^2      
- J_{\alpha+\frac12}J_{\alpha-\frac32}   
- J_{\alpha-\frac12}J_{\alpha+\frac32}   
\right) \left( {\pi\rho(0) x} \right).
\end{equation}
It enjoys the matching condition
\begin{equation}
\lim_{x\rightarrow \infty} \rho_{\rm S}(x)=\rho(0).
\end{equation}

\setcounter{equation}{0}
\section{Conclusion and speculation}
\noindent
In the present work we have proven two theorems
concerning the asymptotic behavior of orthogonal polynomials 
for a broad class of measures. 
They show a remarkable asymptotic universality 
near the origin, a relation previously only known for the classical
orthogonal polynomials of generalized Laguerre and Jacobi type \cite{Sze}. 
Furthermore, we have used this highly universal behavior 
to prove universality
conjectures regarding the spectral densities and their
$s$-point correlators in the microscopic limit.

\vspace{12pt}\noindent
All of these proofs refer to the zero-dimensional matrix model 
language alone. 
And, as we have emphasized in the Introduction, 
the assumption that this universality
can be applied to full $d$-dimensional quantum gauge theories
with fermionic fields is {\em a priori} very far from obvious.
Of course, our matrix model universality theorems are
interesting in their own right. But from a particle physics perspective
the real focus should be on the application of 
these theorems to the Dirac operator spectrum \cite{V93}.

\vspace{12pt}\noindent
How could we imagine the link between random matrix models and gauge
theories established in this connection? 
Consider a Euclidean
four-dimensional 
${\rm SU}(N_c)$ 
gauge theory coupled to $N_f$ light fermions in, say, 
the fundamental representation. Assume that 
$N_f$ is small enough to
allow for the conventional scenario of 
spontaneous chiral symmetry breaking.
Let us simply call such a theory ``QCD''. 
Its partition function can
be written formally as
\begin{equation}
Z_{\rm QCD}=
\int \frac{{\cal D}A_\mu}{\rm (Gauge)}
\prod_{f=1}^{N_f}{\cal D}\bar{\psi}_f{\cal D}\psi_f
\,\e^{-S[A]-\int\bar{\psi}_f ({\rm i} D\!\llap{/}+m_f)\psi_f} ~, 
\end{equation}
where $S[A]$ is the gauge field part of the action. 
Consider the theory in a finite volume $V_4$. 
Let us now seek a
low-energy effective description of this theory.
One way is the chiral Lagrangian ($\chi$L) approach,
which in this case introduces considerable simplification due to the
suppression of all derivative terms 
(since one is interested only in the
very soft modes) \cite{LS}. 
Here the chiral condensate at zero mass,
$\langle \bar{\psi}\psi\rangle \equiv \Sigma$ is by
construction assumed to be non-vanishing.
Below is a schematic picture of how one could imagine 
random matrix theory to fit into this framework:
\[
\begin{array}{ccc}
Z_{\rm QCD}=
\int \frac{{\cal D}A_\mu}{\rm (Gauge)}
\,\e^{-S[A]}\,\prod_f\det ({\rm i} D\!\llap{/}+m_f)
&
\stackrel{\mbox{(b)}}{\longrightarrow}&
Z_{\chi{\rm UE}}=\int \d M\,\e^{-NV(M)}\,\prod_f\det
(M+m_f)\\
\ & \ & \ \\
\downarrow \mbox{(a)}&
\ &
\mbox{(c)}
\downarrow 
\\
\ & \ & \ \\
Z_{\chi {\rm L}}=\int_{{\rm SU}(N_f)} \d U\,
\e^{N\Sigma\,{\rm tr}\,{m} U \, \e^{{\rm i}\frac{\theta}{N_f}}+c.c.}&
\stackrel{\mbox{(d)}}{\longleftarrow} &
Z_{\chi{\rm GUE}}=\int \d M\,\e^{-N\Sigma^2\,{\rm tr}\,M^2}
\,\prod_f \det(M+m_f)
\end{array}
\]

\vspace{12pt}\noindent
This link (a)
is, however, based on the 
global symmetry breaking pattern
${\rm SU}(N_f)_{\rm L}\times{\rm SU}(N_f)_{\rm R}
\rightarrow{\rm SU}(N_f)_{\rm V}$ alone 
and thus by no means
refers to the microscopic theory.

\vspace{12pt}\noindent
Let us now consider a different route (b)-(c)-(d):
suppose we start with QCD in a finite volume, 
and with a sharp ultraviolet cut-off. 
(One would like to think of conventional lattice regularizations,
were it not for the known difficulties of defining massless fermionic
degrees of freedom in that case.)
Consider the formal change of integration variables
from the gauge field $A_{\mu}(x)$ to the Dirac operator
${\rm i} D\!\llap{/}=A\!\llap{/}(x)+{\rm i} \partial\!\llap{/}~
\equiv M$
itself (b).
Since $V_4$ is the size of the matrix $M$, we write $N=V_4$.
Although we can hardly imagine computing the Jacobian $J$ of
this transformation
the resulting partition function would formally look like the
random matrix model above,
with $V(M)=S[A(M)]-\log J(M)$.
Once phrased in a matrix model language, we can apply our Theorem 1
to substitute the uncomputable $V(M)$ by
another simple measure, say that of the Gaussian unitary ensemble (GUE)
with $V(M)=\Sigma^2\,{\rm tr}\,M^2$.
This GUE can be shown to be expressible as
the zero-dimensional chiral Lagrangian 
in the microscopic limit (d) \cite{V93}.
If the relation (b) between
a properly regularized version of QCD
and a particular random matrix model
were not still very formal and weakly established, 
the diagram would otherwise close itself full circle. 
The diagram above at least serves as an intuitive picture 
outlining the logical routes
between the different formulations, and the r\^{o}le played by our
universality theorems proposed in this connection.\\

\vspace{12pt}\noindent
{\large {\bf Acknowledgements}}\\

\noindent
P.H.D.~would like to thank J.~Verbaarschot for an early discussion, 
and P.~Nevai and W.~Van Assche for e-mail correspondences.
G.A.~wishes to thank the Niels Bohr Institute
for its warm hospitality while this work
was being started.
The work of S.N.~is supported by 
JSPS Postdoctoral Fellowships for Research Abroad.

\end{document}